\documentclass{nature}
\usepackage{graphicx}
\usepackage{amssymb}
\usepackage{dcolumn}
\usepackage{bm}
\usepackage{epstopdf}
\usepackage{epsfig}
\usepackage{color}
\usepackage{lineno}
\usepackage[colorlinks=true, letterpaper=true, pdfstartview=FitV, linkcolor=blue, citecolor=blue, urlcolor=blue]{hyperref}

\begin{document}

\title{Multiple Unpinned Dirac Points in Group-Va Single-layers with Phosphorene Structure}

\author{Yunhao Lu$^{1,2,\star}$, Di Zhou$^1$,Guoqing Chang$^{3,4}$,Shan Guan$^5$,Weiguang Chen$^6$,Yinzhu Jiang$^{1,2}$,Jianzhong Jiang$^{1,2}$,Hsin Lin$^{3,4,\star}$,Xue-sen Wang$^4$,Shengyuan A. Yang$^{5,\star}$, Yuan Ping Feng$^4$,Yoshiyuki Kawazoe$^{7,8}$}

\maketitle

\begin{affiliations}
\item School of Materials Science and Engineering, Zhejiang University, Hangzhou 310027, China
\item State Key Laboratory of Silicon Materials, Zhejiang University, Hangzhou 310027, China
\item Centre for Advanced 2D Materials and Graphene Research Centre, National University of Singapore, Singapore 117546, Singapore
\item Department of Physics, National University of Singapore, Singapore 117542, Singapore
\item Research Laboratory for Quantum Materials, Singapore University of Technology and Design, Singapore 487372, Singapore
\item College of Physics and Electronic Engineering, Zhengzhou Normal University, Zhengzhou 450044, China
\item New Industry Creation Hatchery Center, Tohuku University, Sendai, 980-8579, Japan
\item Institute of Thermophysics, Siberian Branch of Russian Academy of Sciences, Novosibirsk 630090, Russia

$^\star$e-mail: luyh@zju.edu.cn;\, nilnish@gmail.com;\, shengyuan\_yang@sutd.edu.sg
\end{affiliations}

\section*{Abstract}

\begin{abstract}
Emergent Dirac fermion states underlie many intriguing properties of graphene, and the search for them constitute one strong motivation to explore two-dimensional (2D) allotropes of other elements. Phosphorene, the ultrathin layers of black phosphorous, has been a subject of intense investigations recently, and it was found that other group-Va elements could also form 2D layers with similar puckered lattice structure. Here, by a close examination of their electronic band structure evolution, we discover two types of Dirac fermion states emerging in the low-energy spectrum.
One pair of (type-I) Dirac points is sitting on high-symmetry lines, while two pairs of (type-II) Dirac points are located at generic $k$-points, with different anisotropic dispersions determined by the reduced symmetries at their locations. Such fully-unpinned (type-II) 2D Dirac points are discovered for the first time. In the absence of spin-orbit coupling, we find that each Dirac node is protected by the sublattice symmetry from gap opening, which is in turn ensured by any one of three point group symmetries. The spin-orbit coupling generally gaps the Dirac nodes, and for the type-I case, this drives the system into a quantum spin Hall insulator phase. We suggest possible ways to realize the unpinned Dirac points in strained phosphorene.
\end{abstract}

\section*{Introduction}
Recent years have witnessed a surge of research interest in the study of Dirac fermions in condensed matter systems, ranging from
graphene and topological insulator surfaces in two-dimensions (2D) to Dirac and Weyl semimetals in 3D,~\cite{Kane2010,Qi2011,Wan2011,Young2012} which possess many intriguing physical properties owing to their relativistic dispersion and chiral nature. Especially, 2D Dirac fermion states have been extensively discussed in honeycomb lattices, commonly shared by group-IVa elements with graphene as the most prominent example,~\cite{Novo2004,Neto2009,Caha2009,Liu2011,Xu2013} for which Dirac points are pinned at the two inequivalent high-symmetry points $K$ and $K'$ of the hexagonal Brillouin zone (BZ), around which the dispersion is linear and isotropic. Later on, 2D Dirac points on high-symmetry lines were also predicted in some nanostructured materials,~\cite{Wang_review} including graphynes~\cite{Malko2012} and rectangular carbon and boron allotropes.~\cite{LCXu2014,XFZhou2014} However, the possibility of 2D Dirac points at generic $k$-points has not been addressed, and such Dirac point has not been found so far.

Meanwhile, the exploration of 2D materials built of group-Va elements (P, As, Sb, and Bi) has just started. Single- and few-layer black phosphorous, known as phosphorene, have been successfully fabricated, and was shown to be semiconducting with a thickness-dependent bandgap and a good mobility up to $\sim$10$^3$cm$^2$/(V$\cdot$s), generating intense interest.~\cite{Li2014,Xia2014,Liu2014,Qiao2014,Rude2014,Liu2015,Kim2015,Woom2015} While 2D allotropes with different lattice structures have been  predicted and analyzed for the other group-Va elements,~\cite{Zhu2014,Zhang2015,GWang2015,Kamal2015} we note that the puckered lattice structure similar to phosphorene has been demonstrated experimentally for Sb~\cite{Wang2006,Bian2012,Stro2012,WXSexp} and Bi~\cite{Naga2004,Kowa2013,Koku2015,Lu2015} (down to single-layer) grown on suitable substrates, and been predicted for As as well.~\cite{Kamal2015} Motivated by these previous experimental and theoretical works, and in view of the ubiquitous presence of the Dirac fermions and
the associated interesting physics, one may wonder: Is it possible to have Dirac fermion states hosted in such 2D puckered lattices?
A simple consideration shows that here any possible Dirac point cannot occur at high-symmetry points. The reason is that each Dirac point at $\bm k$ must have a time reversal (TR) partner at $-\bm k$ with opposite chirality, whereas the BZ of the puckered lattice has a rectangular shape, of which all the high-symmetry points are invariant under TR. Therefore, if Dirac states indeed exist in such systems, they must be of a type distinct from those in graphene.

In this work, we address the above question by investigating the electronic structures of group-Va 2D puckered lattices. We find that Dirac fermion states not only exist, but in fact occur with two different types: one type (referred to as type-I) of (two) Dirac points are located on high-symmetry lines; while the other type (referred to as type-II) of (four) Dirac points are located at generic $k$-points. Depending on their reduced symmetries, dispersions around these points exhibit different anisotropic behaviors. Points of each type can generate or annihilate in pairs of opposite chiralities, accompanying topological phase transitions from a band insulator to a 2D Dirac semimetal phase, and since they are not fixed at high symmetry points, their locations can be moved around in the BZ. Particularly, to our best knowledge, the novel fully-unpinned (type-II) 2D Dirac points are discovered here for the first time. In the absence of spin-orbit coupling (SOC), each Dirac node is protected from gap opening by a sublattice (chiral) symmetry, which can in turn be ensured by {any one} of three point group symmetries. The inclusion of SOC could gap the Dirac nodes, and in the case of type-I nodes, it transforms the system into a quantum spin Hall (QSH) insulator phase. All these properties make the system distinct from graphene and other 2D materials. We further suggest that the novel unpinned Dirac points can be experimentally realized by the strain engineering of phosphorene. Our discovery therefore greatly advances our fundamental understanding of 2D Dirac points, and it also suggests a promising platform for exploring interesting effects with novel types of Dirac fermions.

\section*{Results}

A group-Va pnictogen atom typically forms three covalent bonds with its neighbors.  As shown in Fig.1 for a single-layer phosphorene structure, the P atoms have strong $sp^3$-hybridization character hence the three P-P bonds are more close to a tetrahedral configuration. This results in two atomic planes (marked with red and blue colors) having a vertical separation of $h$ comparable to the bond length. In each atomic plane, the bonding between atoms forms zigzag chains along $y$-direction. The unit cell has a 4-atom basis, which we label as $A_U$, $B_U$, $A_L$, and $B_L$ (see Fig.1(c)), where $U$ and $L$ refer to the upper- and lower-plane respectively.
The structure has a nonsymmorphic $D_{2h}(7)$ space group which includes the following elements that will be important in our discussion: an inversion center $i$, a vertical mirror plane $\sigma_v$ perpendicular to $\hat{y}$, and two 2-fold rotational axes $c_{2y}$ and $c_{2z}$. Note that due to the puckering of the layer, the mirror planes perpendicular to $\hat{x}$ and $\hat{z}$ are broken.
With the same valence electron configuration, As, Sb, and Bi possess allotropes with similar puckered lattice structures.

To study the electronic properties, we performed first-principles calculations based on the density functional theory (DFT). The details are described in the Methods. The calculated geometric parameters of group-Va 2D puckered lattices with $D_{2h}(7)$ symmetry are summarized in the Supplementary Information. The obtained structures agree with the experiments and other theoretical calculations.~\cite{Qiao2014,Lu2015,WXSexp,Kamal2015} The lattice constants $a>b$, reflecting that the inter-chain coupling is weaker than the coupling along the zigzag chains. The angle $\theta_2$ increases from $\sim70^\circ$ for P to $\sim85^\circ$ for Bi whereas $\theta_1$ remains $\sim95^\circ$. The inter-plane separation $h$ as well as the bond lengths $R_1$ and $R_3$ increase by almost 1~\textup{\AA}; from P to Bi, while $R_2$, the distance between sites of neighboring zig-zag chains, increases only slightly, implying that the inter-chain coupling becomes relatively more important with increasing atomic number.

We first examine their corresponding band structures without SOC, whose effect will be discussed later. The results are shown in Fig.2. The puckered lattice of P is a semiconductor with a bandgap around $\Gamma$-point. From P to Bi, the direct bandgap at $\Gamma$-point keeps decreasing, and a drastic change occurs from Sb to Bi where linear band crossings can be clearly spotted along the $\Gamma$-$X_2$ line. Examination of the band dispersion around the two points (labeled as $D$ and $D'$ in Fig.1(d)) shows that they are indeed Dirac points (see Fig.3(a)). Furthermore, along $\Gamma$-$X_1$ line, there gradually appear two sharp local band extremum points for both conduction and valence bands, where the local gap decreases from P to Sb with the two bands almost touching for Sb, yet the trend breaks for Bi. Remarkably, close examination reveals that for Sb and Bi, close to each extremum point there are actually two Dirac points on the two sides of the $\Gamma$-$X_1$ line (see Figs.1(d) and 3(b)). The energy dispersions around these Dirac points are shown in Fig.1(e) and 1(f), clearly demonstrating the Dirac cone characters. Therefore, two types of Dirac points with distinct symmetry characters exist in this system: one pair of type-I Dirac points ($D$ and $D'$) sitting on high-symmetry lines and two pairs of type-II Dirac points (near $F$ and $F'$) at generic $k$-points.

The band evolution around $\Gamma$-point from Sb to Bi and the appearance of type-I Dirac points in Fig.2 are reminiscent of a band-inversion process. Indeed, by checking the parity eigenvalues at $\Gamma$, one confirms that the band order is reversed for Bi around $\Gamma$-point (see Supplementary Information). For a better understanding, we
construct a tight-binding model trying to capture the physics around $\Gamma$-point. Since the low-energy bands there are dominated with $p_z$-orbital character (Fig.2), we take one orbital per site, and include couplings along $R_1$ and $R_2$ in the same atomic plane (with amplitudes $t_1$ and $t_2$ respectively) as well as nearest-neighbor inter-plane hopping along $R_3$ (with amplitude $t_\bot$) (see Supplementary Information). Written in the basis of $(A_U, A_L, B_U, B_L)$, the Hamiltonian takes the form:
\begin{equation}\label{H}
\mathcal{H}(\bm k)=\left[
                     \begin{array}{cc}
                       0 & Q(\bm k) \\
                       Q^\dagger(\bm k) & 0 \\
                     \end{array}
                   \right],
\end{equation}
where $Q(\bm k)$ is a 2$\times$2 matrix of the Fourier transformed hopping terms (see Supplementary Information). The Hamiltonian (\ref{H}) can be diagonalized and possible band crossings can be probed by searching for the zero-energy modes, which exist when the condition $\lambda\equiv t_\bot/[2(t_1+t_2)]<1$ is satisfied, with two band touching points at $(0, \pm k_D)$ where $k_D=(2/b)\arccos(\lambda)$. The direct gap at $\Gamma$ can be obtained as $\Delta=2[t_\bot-2(t_1+t_2)]$. Hence this simple model indeed captures the emergence of two Dirac points $D$ and $D'$, along with a transition as parameter $\lambda$ varies: when $\lambda>1$, the system is a band insulator; when $\lambda<1$, it is a 2D Dirac semimetal. The transition occurs at the critical value $\lambda_c=1$ when the conduction and valence bands touch at $\Gamma$-point and the band order starts to be inverted. {This corresponds to a quantum (and topological) phase transition,~\cite{Volovik} during which there is no symmetry change of the system.}

Model (\ref{H}) captures the trend observed in DFT results. The overlap between-$p_z$ orbitals is larger along the $R_3$ bond, hence one expects that $t_\bot>t_1>t_2$. {By fitting the DFT bands around $\Gamma$-point, one finds that from P to Bi, $t_\bot$ decreases a lot, while $t_2$ increases and becomes relatively more important (Supplementary Information).} The result shows that $(t_\bot,t_1,t_2)$ changes from $(2.50,0.77,0.33)$ for Sb to $(1.86,0.63,0.35)$ for Bi (units in eV).
Hence $\lambda$ crosses the critical value from Sb to Bi, indicating the band inversion at $\Gamma$ and the appearance of two Dirac points.

The emergence of low-energy relativistic chiral modes is the most remarkable property of Dirac points.~\cite{Volovik} To explicitly demonstrate this, we expand Hamiltonian (\ref{H}) around each Dirac point, which leads to the low-energy Hamiltonian
\begin{equation}\label{Heff}
H_{\tau}(\bm q)=v_x q_x\sigma_y+\tau v_y q_y\sigma_x,
\end{equation}
where $\bm q$ is the wave-vector measured from each Dirac point, $\tau=\pm 1$ for $D$ and $D'$, $\sigma_i$'s are Pauli matrices for the sub-space spanned by the two eigenstates at the Dirac point (apart from the Bloch phase factor): $|u_1\rangle=(0,0,1,-1)/\sqrt{2}$ and $|u_2\rangle=(1,-1,0,0)/\sqrt{2}$, and $v_x=at_\bot (t_1-t_2)/(t_1+t_2)$ and $v_y=b\sqrt{4(t_1+t_2)^2-t_\bot^2}$ are the two Fermi velocities. The form of (\ref{Heff}) may also be argued solely from symmetry.
Compared with graphene, these type-I points are unpinned from the high-symmetry points. They can be shifted along $\Gamma$-$X_2$ (and even pair-annihilated) by varying system parameters such as $\lambda$, although they cannot go off the line as constrained by the symmetries. In addition, different from graphene,~\cite{Neto2009} the dispersion here is anisotropic, characterized by two different Fermi velocities.

Next, we turn to the fully-unpinned type-II Dirac points. The four type-II Dirac points start to appear for Sb in our DFT result, located close to the $\Gamma$-$X_1$ line. They can be more clearly seen for Bi (see Fig.3(b)). Again the band evolution implies a local band inversion near $F$ and $F'$. Here $F$ and $F'$ (on $\Gamma$-$X_1$) are the mid-points of the lines connecting each pair of the type-II points. The low-energy bands there are mainly of $p_x$-orbital character. To reproduce the fine features using a tight-binding model would require more hopping terms. Instead, we construct a low-energy effective Hamiltonian around point $F$ ($F'$) based on symmetry analysis. There the Hamiltonian is constrained by $\sigma_v$ which maps inside each pair (labeled by $\mu=\pm 1$ for $F$ and $F'$), and by $i$, $c_{2y}$, $c_{2z}$, and TR that map between the two pairs. Expansion to leading order in each wave-vector component $q_i$ gives (see Supplementary Information)
\begin{equation}\label{H2}
H_\mu(\bm q)=w q_x\sigma_y+(-m_0+\mu w' q_x+m_1 q_y^2)\sigma_x,
\end{equation}
where $\bm q$ is measured from $F$ (or $F'$), $w$, $w'$, $m_0$, and $m_1$ are expansion coefficients. Two Dirac points appear at $(0,\pm q_0)$ with $q_0=\sqrt{m_0/m_1}$ when $\emph{sgn}(m_0/m_1)=1$, corresponding to a local band inversion around $\bm q=0$. Further expansion of the Hamiltonian around the Dirac point $(0,\nu q_0)$ ($\nu=\pm 1$) leads to
\begin{equation}
\tilde{H}_\mu^\nu =w q_x \sigma_y+[2\nu q_0(q_y-\nu q_0)+\mu w'q_x]\sigma_x.
\end{equation}
This demonstrates that the two points at $\nu=\pm 1$ are of opposite chirality, as required by $\sigma_v$. The dispersion is highly anisotropic (at leading order, characterized by three parameters: $w$, $q_0$, and $w'$) because the Dirac point is at a generic $k$-point with less symmetry constraint, as compared with type-I Dirac points.

Unlike in 3D systems, Dirac nodes in 2D have a codimension of two hence are generally not protected from gap-opening.~\cite{Volovik} In the absence of SOC, however, the Dirac nodes here are stable due to the protection by sublattice (chiral) symmetry between $\{A_i\}$ and $\{B_i\}$ ($i=U,L$) sites, which allows the definition of a winding number~\cite{Schn2008,Yang2014} (i.e., quantized Berry phase in units of $\pi$) along a closed loop $\ell$ encircling each Dirac point: $N_\ell=\oint_\ell \mathcal{A}_{\bm k}\cdot\emph{d}\bm k/\pi=\pm 1$, where $\mathcal{A}_{\bm k}$ is the Berry connection of the occupied valence bands. And for a 2D Dirac point, the sign of $N_\ell$ (or the $\pm \pi$ Berry phase) is also referred to as the chirality.~\cite{Neto2009} Using DFT results, we numerically calculate the Berry phase for each Dirac point and indeed confirm that they are quantized as $\pm \pi$. The signs are indicated in Fig.1(d).

More interestingly, in the puckered lattice with a four-atom basis in a non-coplanar geometry, the sublattice symmetry can be ensured by \emph{any one} of three independent point group symmetries: $i$, $c_{2y}$, and $c_{2z}$. The resulting protection of Dirac nodes can be explicitly demonstrated in low-energy models. For example, consider the type-I points described by Eq.(\ref{Heff}).
There the representations of $i$, $c_{2y}$, and $c_{2z}$ (denoted by $\mathcal{P}$, $\mathcal{R}_y$, and $\mathcal{R}_z$, respectively) are the same, viz., $\sigma_x$.
Then the symmetry requirement $\mathcal{R}_yH_\tau(q_x,q_y)\mathcal{R}_y^{-1}=H_\tau(-q_x,q_y)$ by $c_{2y}$ directly forbids the presence of a mass term $m\sigma_z$.
Meanwhile, since $i$ and $c_{2z}$ maps one valley to the other, they protect the Dirac nodes when combined with TR (\emph{or} $\sigma_v$ if it is unbroken), e.g., considering the combined symmetry of $c_{2z}$ and TR (with representation $\mathcal{T}=K$ the complex conjugation operator): $(\mathcal{R}_z\mathcal{T})H_\tau(\bm q)(\mathcal{R}_z\mathcal{T})^{-1}=H_\tau(\bm q)$, which again forbids a mass generation.
The underlying reason $i$, $c_{2y}$, and $c_{2z}$ each protects the Dirac node is that they each maps between the two sublattices hence ensures the sublattice (chiral) symmetry. In comparison, the mirror plane $\sigma_v$ maps inside each sublattice, hence it alone cannot provide such protection. This reasoning is general and applies to type-II points as well. (In model (\ref{H2}), $i$, $c_{2y}$, and $c_{2z}$ have representations as $\sigma_x$ by construction, and when combined with TR, again each forbids the generation of a mass term $\sim m\sigma_z$. See Supplementary Information.)
We stress that the three symmetries $i$,  $c_{2y}$, and $c_{2z}$ \emph{each} protects the Dirac points independent of the other two. For example, we could disturb the system as in Fig.4 such that only one of the three symmetries survives. The corresponding DFT results confirm that the Dirac nodes still exist. Thus the crystalline symmetries actually offer multiple protections for the Dirac nodes in the current system.

SOC could break the sublattice symmetry. Hence when SOC is included, the Dirac nodes would generally be gapped.~\cite{YK2015} For type-I points, treating SOC as a perturbation, its leading-order symmetry-allowed form is
$H_\mathrm{SOC}=\tau\Delta\sigma_z s_z$,
where $s_z$ is Pauli matrix for real spin. This is similar to the intrinsic SOC term in graphene,~\cite{Kane2005a} which opens a gap of $2|\Delta|$ at the Dirac points. For the type-II points, we obtain $H_\mathrm{SOC}=\eta q_y\sigma_z s_z$ in model (\ref{H2}) hence a gap of $2q_0|\eta|$ is also opened at these Dirac points. Gap opening by SOC is closely related to the QSH insulator phase.~\cite{Kane2005a,Kane2010,Qi2011} Here the band topology can be directly deduced from the parity analysis at the four TR invariant momenta.~\cite{Fu2007} This means that only the band inversion at $\Gamma$ between the two type-I points contributes to a nontrivial $\mathbb{Z}_2$ invariant; whereas that associated with type-II points does not. It follows that Sb is topologically trivial since it has only type-II Dirac points, while Bi is nontrivial since it has additional type-I points. These results are in agreement with previous studies.~\cite{Lu2015}

Breaking all three symmetries $i$, $c_{2y}$ and $c_{2z}$ can also generate a trivial gap term $m\sigma_z$ at the Dirac points, which competes with the SOC gap. For example, this happens when each atomic-plane forms additional buckling structure.~\cite{Lu2015} Nevertheless, as long as the trivial mass term does not close the SOC-induced gap, by adiabatic continuation the band topology will not change.

\section*{Discussion}

Due to their different locations and the associated symmetries, the two types of Dirac points here exhibit properties distinct from that of graphene. With preserved sublattice symmetry and in the absence of SOC, the Dirac nodes are topologically stable---they can only disappear by pair-annihilation between opposite chiralities. This is unlikely for graphene since the Dirac points there are pinned at the high-symmetry points. In contrast, the two types of Dirac points here are less constrained. Pair-annihilation (pair-generation) indeed occurs during the quantum phase transition as observed from the band evolution.

It is noted that similar type-I points were also predicted in a few nanostructured materials.~\cite{Malko2012,LCXu2014,XFZhou2014} Meanwhile the type-II points discovered here are completely new. They are fully-unpinned and has highly anisotropic dispersions. With this discovery, now we can have an almost complete picture: 2D Dirac points can occur at high-symmetry points, along high-symmetry lines, and also at generic $k$-points.

It is possible to have Dirac points, originally sitting at high-symmetry points, to become unpinned when crystalline symmetry is reduced due to structural distortions. However, we stress that the type-II points here are distinct in that they are realized in a native crystalline structure with relatively high symmetry. Only in such a case, we can have a sharp contrast between generic $k$-points where the group of wave vectors is trivial and the high-symmetry $k$-points where the group is non-trivial, and accordingly the type-II point can move around (hence fully-unpinned) without any symmetry-breaking. More importantly, it is just because that type-II points occur in a state with high symmetry that the Dirac nodes can be protected (in the absence of SOC): as we discussed, the various crystalline symmetries ensure the protection of the Dirac nodes from gap opening.

It is remarkable that the two different types of Dirac points can coexist in the same 2D material.
We emphasize that it is a result of the lattice structure and the valence character of the pnictogen elements. Our DFT result indeed shows that even starting from the P lattice, the two types of Dirac points can be separately tuned to appear or disappear by lattice deformations. For example, we find that the type-II Dirac points can be generated in phosphorene by applying uniaxial tensile strains along the $y$-direction. The DFT result
in Fig.5 and Fig.6 indeed shows the band inversion on $\Gamma$-$X_1$ and the formation of four type-II Dirac
points around a strain of 16\%. Since phosphorene has excellent mechanical properties and a critical strains $>25$\% has been predicted,~\cite{strain2} it is promising that the novel strain-induced topological phase transitions and the appearance of type-II Dirac points can be directly observed in strained phosphorene. The lattice deformation that produces type-I points is discussed in the Supplementary Information. Similar scenarios occur for other group-Va elements as well.

So far, single-layer As, Sb, and Bi in their free-standing form have not been realized yet. Nevertheless, in view of the rapid progress in experimental techniques, we expect that these materials could be fabricated in the near future. Especially, for Sb and Bi, the puckered structures have been demonstrated by PVD growth down to single-layer thickness on suitable substrates.~\cite{Wang2006,Bian2012,Stro2012,WXSexp,Naga2004,Kowa2013,Koku2015,Lu2015} Besides the topological properties, the presence of Dirac states is expected to endow these 2D materials with many intriguing properties for applications, such as the very high mobility, the half-quantized quantum Hall effect,~\cite{QHE} the universal optical absorption~\cite{optic} and etc. Due to the highly anisotropic dispersions of these new Dirac points, the electronic transport properties such as the conductivities would show strong direction dependence.  In addition, since there is no symmetry connection between the two types of Dirac points, when they are both present, it is possible to independently shift each type of points relative to the Fermi level, e.g. by strain engineering, leading to self-doping and even the interesting scenario with both electron-like and hole-like Dirac fermions in the same system. With the multiple Dirac points with different chiralities, it is possible to further control the carriers near different Dirac points for valleytronic applications.

In conclusion, based on first-principles calculations of 2D allotropes of group-Va elements with puckered lattice structure, we predict the coexistence of two different types of Dirac points: Dirac points on high symmetry lines and at generic $k$-points. In particular, the 2D Dirac points at generic $k$-points are fully-unpinned, have highly anisotropic dispersions, and are discovered here for the first time. Combined with low-energy effective modeling, we unveil the low-energy properties of these Dirac points. We show that their appearance is associated with the band inversion process corresponding to a topological phase transition. The topology/symmetry protection of the Dirac nodes is analyzed in detail. Interestingly, because of the unique lattice structure, there is a triple-protection of the nodes by three independent point group symmetries. This also implies versatile methods to control the locations as well as the dispersions around the Dirac points. When SOC is strong, the Dirac nodes are gapped, and in the case of type-I points (such as for Bi), this drives the system into a QSH insulator phase. We further show that the topological phase transition and the novel unpinned Dirac points can be realized in strained phosphorene. Our work represents a significant conceptual advance in our fundamental understanding of 2D Dirac points. The result also suggests a new platform to explore novel types of 2D Dirac fermions both for their fascinating fundamental properties and for their promising electronic and valleytronic applications.

\section*{Methods}
\subsection{First-principles calculations.}
Our first-principles calculations are based on the density functional theory (DFT) implemented in the Vienna ab-initio simulation package.~\cite{vasp1} The projector augmented wave pseudopotential method is employed to model ionic potentials.~\cite{Bloc1994} Kinetic energy cutoff is set to 400 eV and $k$-point sampling on the rectangular BZ is with a mesh size 20$\times$20. The minimum vacuum layer thickness is greater than 20 \textup{\AA}; which is large enough to avoid artificial interactions with system images. The structure optimization process is performed including SOC with the local density approximation (LDA) for the exchange-correlation energy\cite{Solo1994} and with van der Waals corrections in the Grimme implementation.~\cite{Grim2006} The force convergence criteria is set to be 0.01 eV/\textup{\AA}. Hybrid functional (HSE06)\cite{hse1} is used for the band structure calculations.

\bibliographystyle{apsrev4-1}

\begin{addendum}
\item [Acknowledgements]
The authors thank D.L. Deng and Shengli Zhang
for helpful discussions. This work was supported by NSFC (Grant No. 11374009, 61574123, and 21373184), National Key Basic
Research Program of China (2012CB825700), SUTD-SRG-EPD2013062, Singapore MOE Academic Research Fund Tier 1 (SUTD-T1-2015004), A*STAR SERC 122-PSF-0017, and AcRF R-144-000-310-112. H.L. acknowledges support by Singapore National Research Foundation under NRF Award No. NRF-NRFF2013-03. The authors
gratefully acknowledge support from SR16000 supercomputing resources
of the Center for Computational Materials Science, Tohoku University.

\item [Author Contributions]
Y.L. and D.Z. performed the first-principles calculations. G.C. and S.G. helped with the data analysis and model fitting. H.L. and S.A.Y. did the analytical modeling and symmetry/topology analysis. W.C., Y.J., J.J., and X.W. participated in the discussion and analysis. X.W., Y.P.F., and Y.K. supervised the work. Y.L., S.G., and S.A.Y. prepared the manuscript. All authors reviewed the manuscript.

\item [Competing Interests]
The authors declare no competing financial interests.

\item [Correspondence]
Correspondence should be addressed to Yuhao Lu, Hsin Lin or Shengyuan A. Yang.

\item [Additional Information]
Supplementary information is available in the online version of the paper.

\end{addendum}

\newpage
\begin{figure}
  \begin{center}\label{fig1}
   \epsfig{file=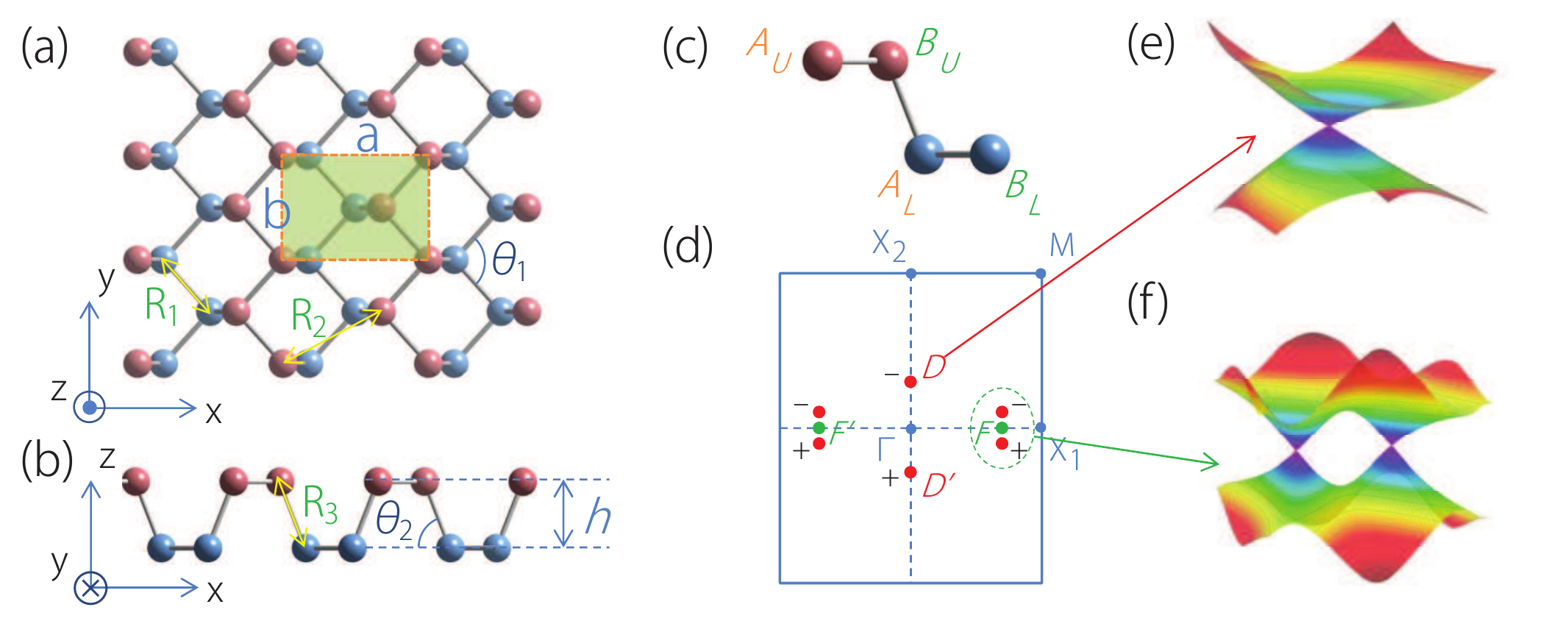,width=16cm}
  \end{center}
  \caption{(a,b) Top- and side-view of the 2D puckered lattice structure. The green shaded region marks the unit cell. (c) Four-atom basis sites in a unit cell. (d) 2D Brillouin zone with high-symmetry points. The locations of Dirac points are schematically marked by the red dots: two type-I Dirac points at $D$ and $D'$ on $\Gamma$-$X_2$; and four type-II Dirac points around $\Gamma$-$X_1$ forming two mirror image pairs. $F$ ($F'$) on $\Gamma$-$X_1$ labels the mid-point of each pair. $+$($-$) indicates the chirality of each point. (e) and (f) show the schematic energy dispersions around point $D$ and point $F$ respectively, corresponding to the result in Fig.3.}
\end{figure}

\newpage
\begin{figure}
  \begin{center}\label{fig2}
   \epsfig{file=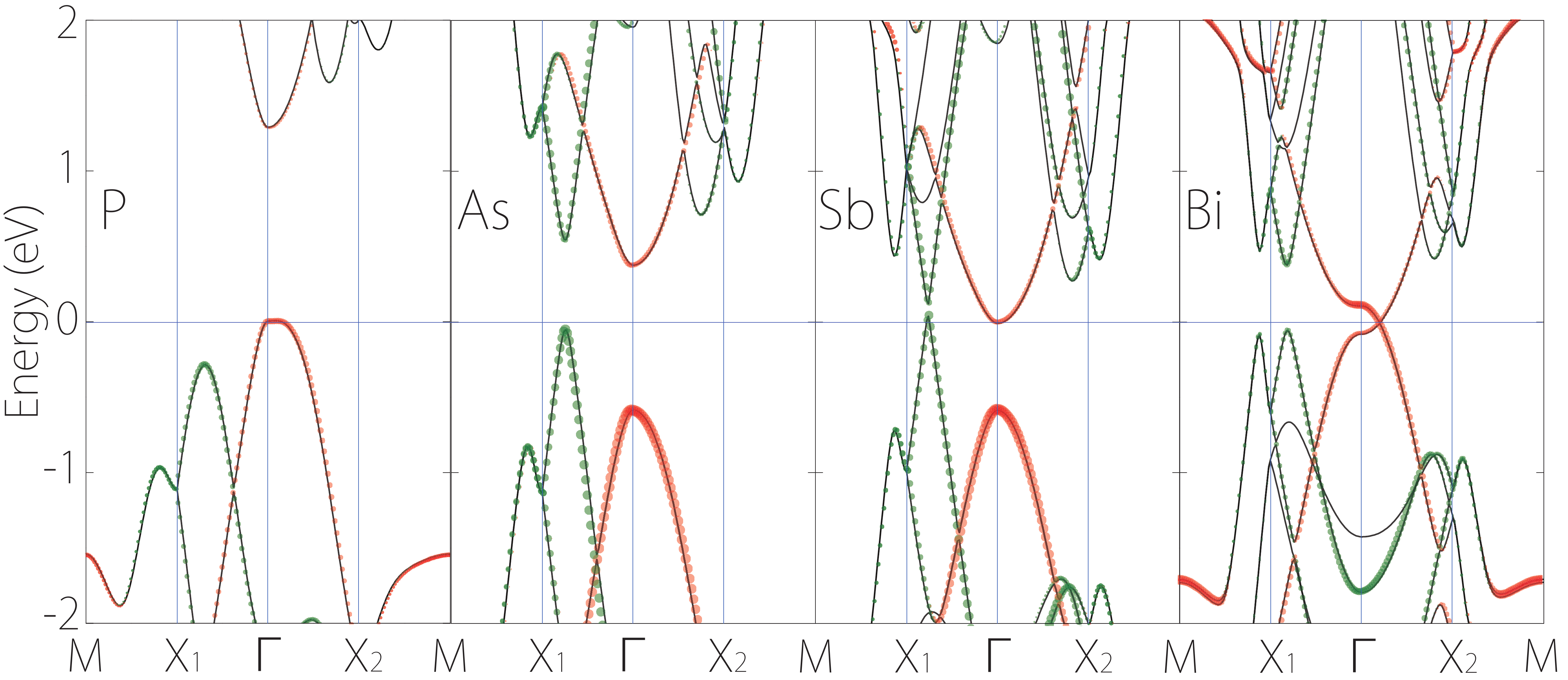,width=16cm}
  \end{center}
  \caption{Band structures of group-Va elements with 2D phosphorene lattice structures in the absence of SOC. The size of red (green) dots denotes the weight of projection onto $p_z$ ($p_x$) atomic orbitals.}
\end{figure}

\newpage
\begin{figure}
  \begin{center}\label{fig3}
   \epsfig{file=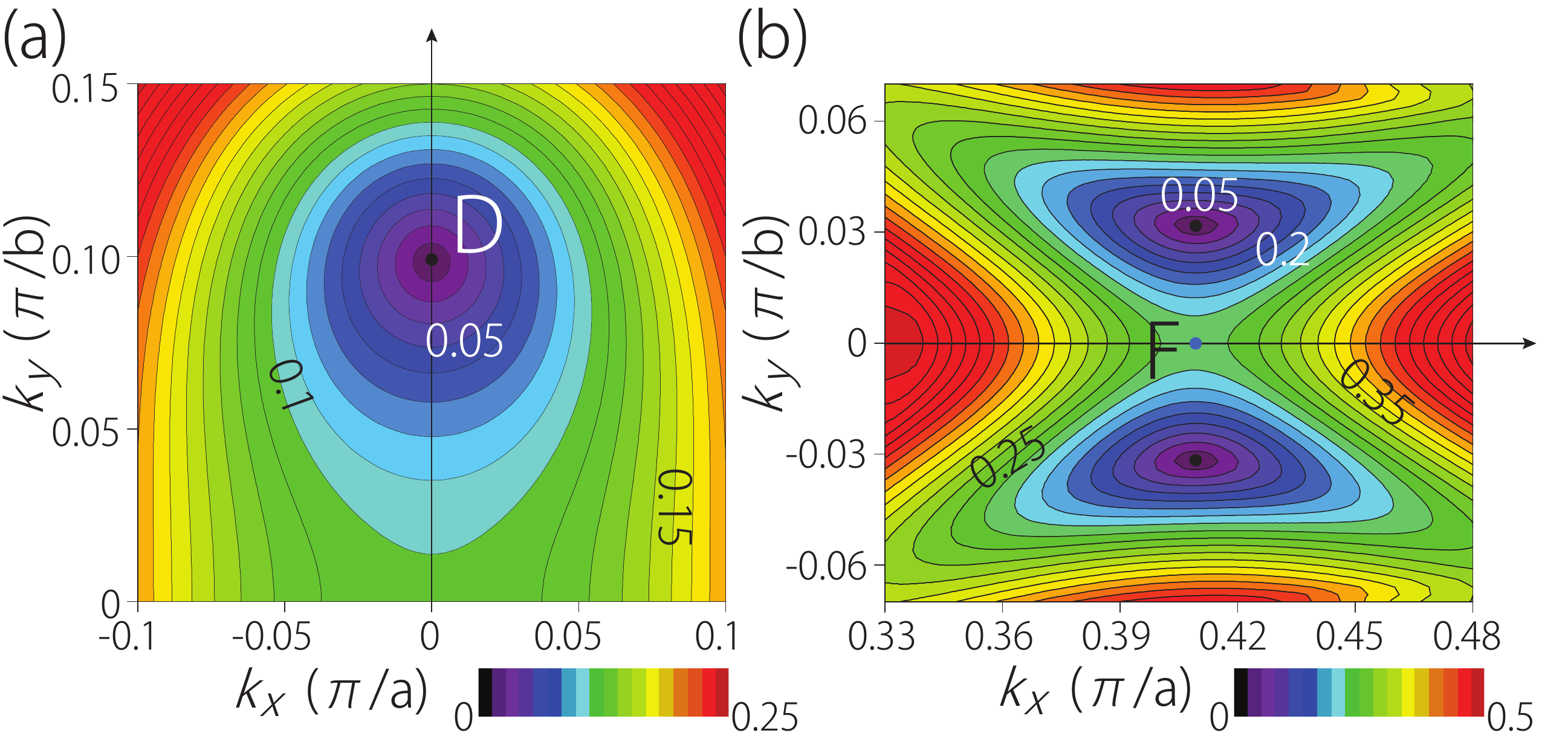,width=16cm}
  \end{center}
  \caption{Dispersion around (a) type-I and (b) type-II Dirac points for 2D puckered Bi. Black dots mark the Dirac point locations. Energy is in unit of eV.}
\end{figure}

\newpage
\begin{figure}
  \begin{center}\label{fig4}
   \epsfig{file=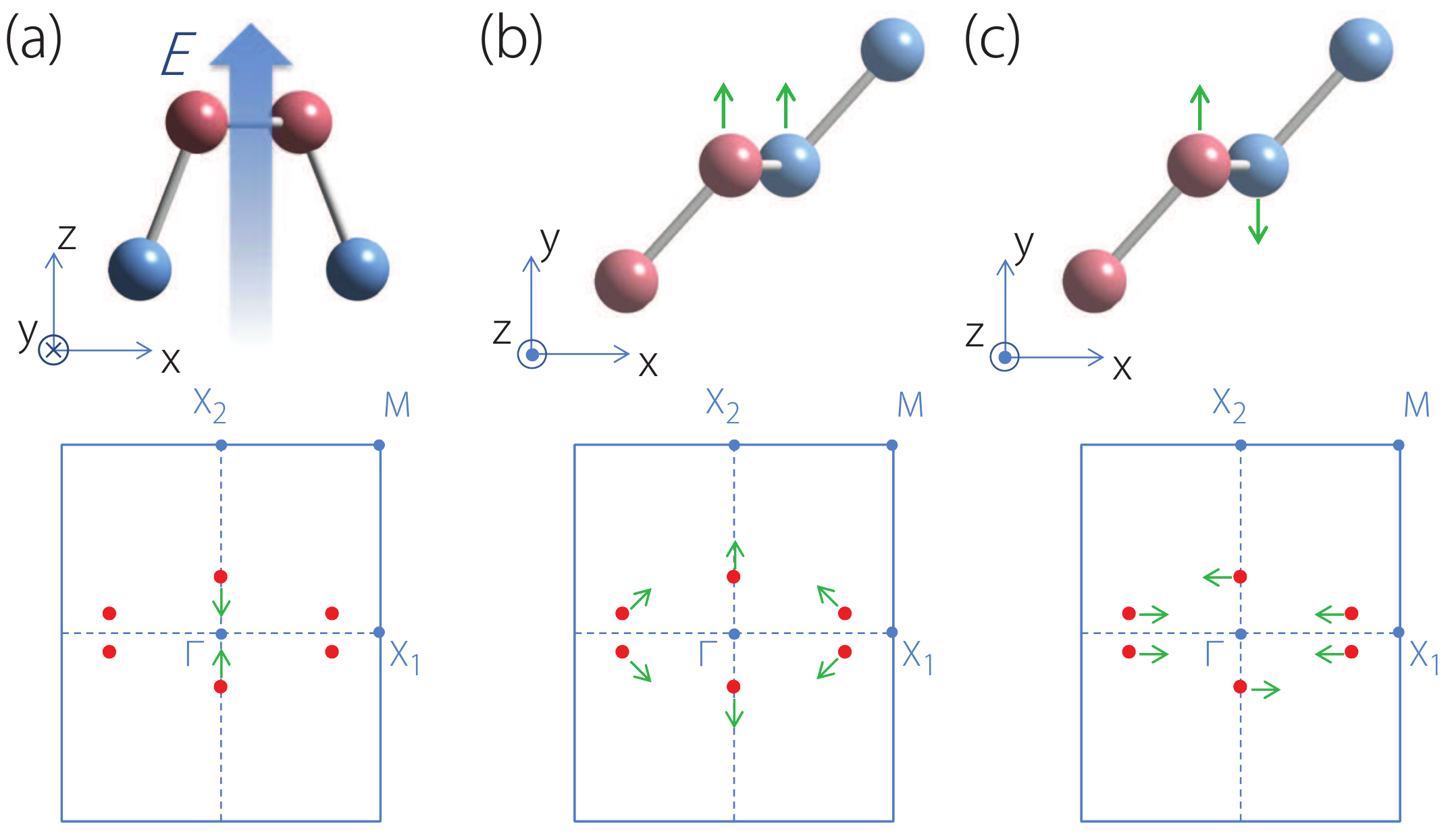,width=16cm}
  \end{center}
  \caption{Perturbations which leave only one of the three symmetries preserved: (a) $c_{2z}$ survives when applying $E$ field in $z$-direction; (b) $c_{2y}$ survives when shifting the two basis sites along $\hat{y}$ in the same direction; (c) $i$ survives when shifting the two sites along $\hat y$ but in opposite directions. For each case, the Dirac nodes are still protected in the absence of SOC. The lower panel in each sub-figure schematically indicate the movement of the Dirac points under each perturbation.  For the case in (a), the type-II points do not show appreciable change in their locations under a weak electric field, while the two type-I points annihilate with each other when they meet at $\Gamma$-point.}
\end{figure}

\newpage
\begin{figure}
  \begin{center}\label{fig5}
   \epsfig{file=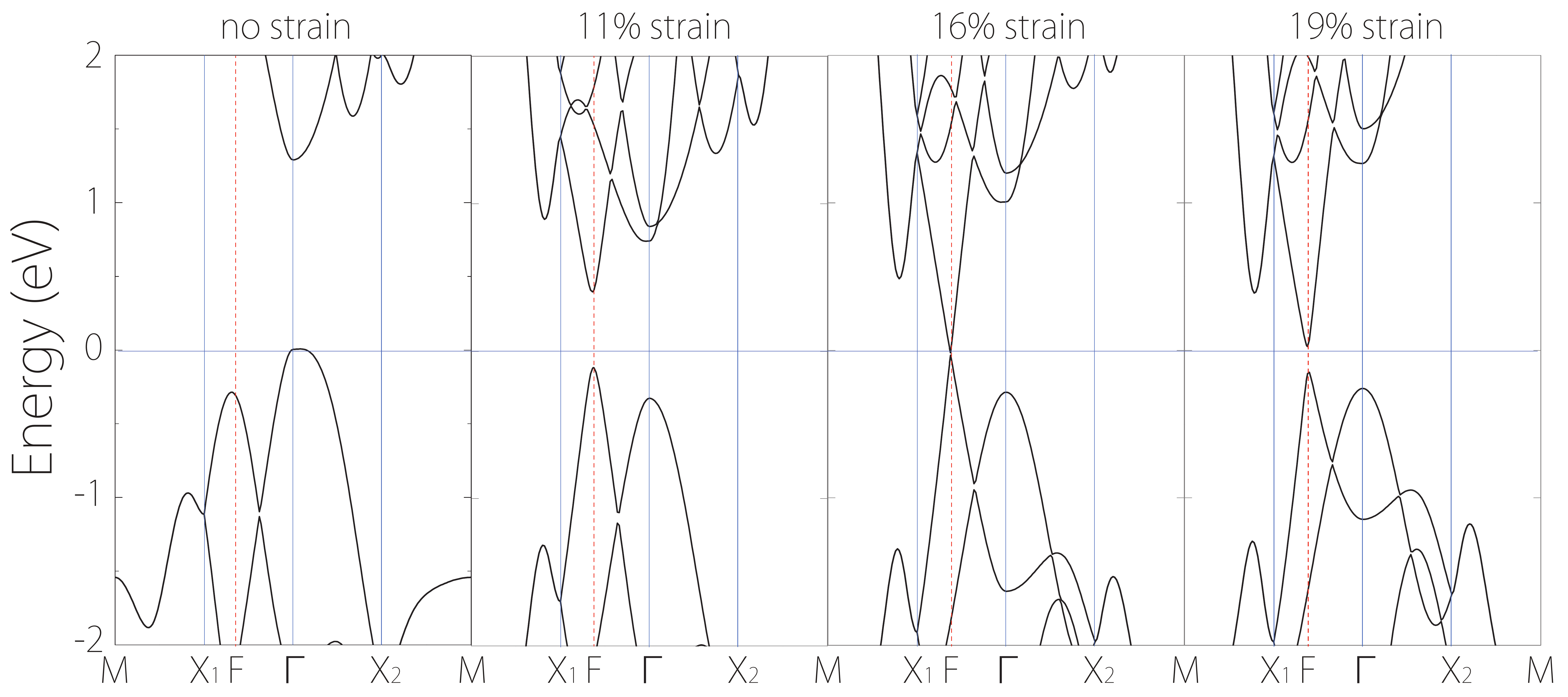,width=16cm}
  \end{center}
  \caption{Band structure of phosphorene with uniaxial tensile strain applied along $y$-direction. Band inversion along $\Gamma$-$X_1$ line occurs around a value of $16\%$.}
\end{figure}

\newpage
\begin{figure}
  \begin{center}\label{fig6}
   \epsfig{file=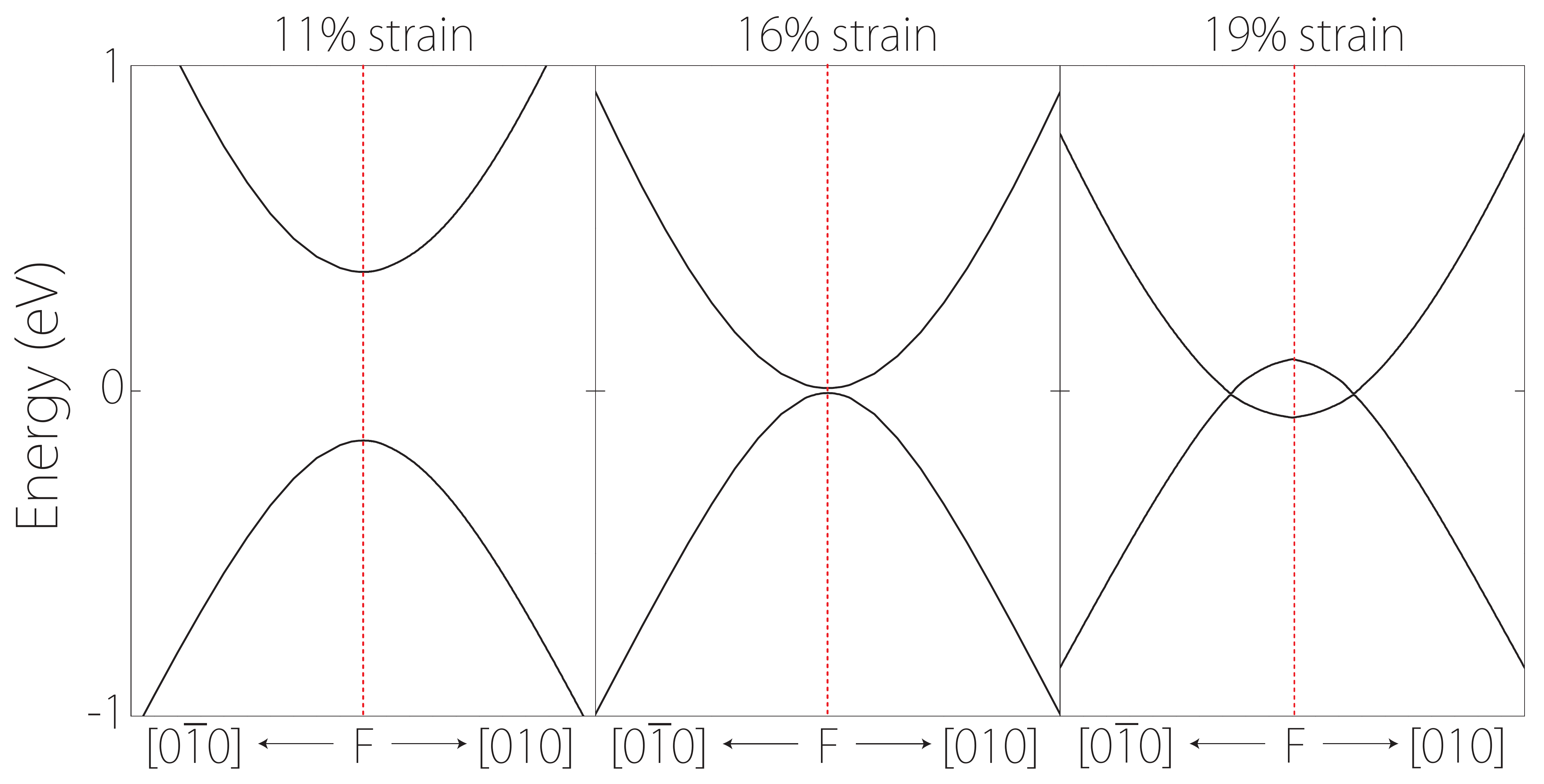,width=16cm}
  \end{center}
  \caption{Band structure of phosphorene with uniaxial tensile strain applied along $y$-direction. Here the dispersion is plotted along a path crossing point $F$ (as indicated in Fig.5) and perpendicular to the $\Gamma$-$X_1$ direction. Two type-II Dirac points near $F$ can be clearly observed at strain value $>16\%$.}
\end{figure}
\end{document}